\documentclass[aip,rsi,reprint]{revtex4-1}
\usepackage{graphicx}

\usepackage[utf8x]{inputenc}
\usepackage{units}
\usepackage{amsmath}
\usepackage{hyperref}  
\hypersetup{colorlinks=true,breaklinks=true,linkcolor=blue,menucolor=blue,pagecolor=blue,urlcolor=blue}
\usepackage{float}
\usepackage{amssymb}
\usepackage{tabularx}

\providecommand{\abs}[1]{\left\lvert#1\right\rvert}
\providecommand{\startt}[1]{\texttt{*}}
\providecommand{\mean}[1]{\left\langle{#1}\right\rangle}
\providecommand{\affDKFZ}{German Cancer Research Center (DKFZ), Biophysics of Macromolecules (B040), Im Neuenheimer Feld 580, D-69120 Heidelberg, Germany}
\providecommand{\affHU}{University of Debrecen, Medical and Health Science Center, Faculty of Medicine, Department of Biophysics and Cell Biology, H-4032 Debrecen, Nagyerdei krt. 98, Hungary}
\providecommand{\multitau}{\mbox{multi-$\tau$}\ }
\providecommand{\mtau}{\multitau}
\providecommand{\Mtau}{\mbox{Multi-$\tau$}\ }
\newcommand{\figref}[1]{\figurename~\ref{#1}}
\newcommand{\Figref}[1]{FIG.~\ref{#1}}
\newcommand{\tabref}[1]{TAB.~\ref{#1}}
\newcommand{\lcite}[1]{Ref.~\onlinecite{#1}}
\renewcommand{\eqref}[1]{Eq.~\ref{#1}}

\begin{document}
	\title{FPGA implementation of a 32x32 autocorrelator array for analysis of fast image series}
	\author{Jan \surname{Buchholz}}
	\author{Jan Wolfgang \surname{Krieger}}
	\affiliation{\affDKFZ}
	\author{G\'{a}bor Mocs\'{a}r}
	\affiliation{\affHU}
	\author{Bal\'{a}zs Kreith}
	\affiliation{\affHU}
	\author{Edoardo \surname{Charbon}}
	\affiliation{Technische Universiteit Delft, Mekelweg 4, 2628 CD Delft, The Netherlands}
	\author{Gy\"{o}rgy V\'{a}mosi}
	\affiliation{\affHU}
	\author{Udo \surname{Kebschull}}
	\affiliation{Goethe-Universit\"{a}t Frankfurt, Senckenberganlage 31, D-60325 Frankfurt, Germany}
	\author{J\"{o}rg \surname{Langowski}}
	\email[e-mail: ]{jl@dkfz.de}
	\homepage[homepage: ]{http://www.dkfz.de/Macromol}
	\affiliation{\affDKFZ}
	
	\date{\today}
	\pacs{42.50.Ar, 42.62.Fi, 78.47.je}
  \keywords{Correlator; FPGA; SPIM; SPAD; FCS; Photon counting and statistics}

	\begin{abstract}
		With the evolving technology in CMOS integration, new classes of 2D-imaging detectors have recently become available. In particular, single photon avalanche diode (SPAD) arrays allow detection of single photons at high acquisition rates ($\ge\unit[100]{kfps}$), which is about two orders of magnitude higher than with currently available cameras. Here we demonstrate the use of a SPAD array for imaging fluorescence correlation spectroscopy (imFCS), a tool to create 2D maps of the dynamics of fluorescent molecules inside living cells. Time-dependent fluorescence fluctuations, due to fluorophores entering and leaving the observed pixels, are evaluated by means of autocorrelation analysis. 
		The \multitau correlation algorithm is an appropriate choice, as it does not rely on the full data set to be held in memory. Thus, this algorithm can be efficiently implemented in custom logic. We describe a new implementation for massively parallel \multitau correlation hardware. Our current implementation can calculate 1024 correlation functions at a resolution of $\unit[10]{\mu s}$ in real-time and therefore correlate real-time image streams from high speed single photon cameras with thousands of pixels. 
	\end{abstract}
	\maketitle

	\section{Introduction}
	Fluorescence correlation spectroscopy\cite{MAGDE1974,MAGDE1974a} (FCS) is a powerful experimental technique for measuring the dynamics of fluorescently labeled molecules in solution and also inside living cells. It allows one to determine the particle number, the diffusion coefficient, flow speeds and also photophysical and chemical reaction rates (for an overview, see Ref.~\onlinecite{KRICHEVS2002}). In FCS the time trace of the fluorescence intensity fluctuations $I(t)$ inside a small observation volume (usually around $\unit[10^{-15}]{l}=1\unit[]{\mu m^3}$) is monitored. The fluctuations originate from particles entering and leaving the focus, or transitions between states having different quantum yields. Faster dynamics of the fluorescing particles also lead to faster fluctuations, which can be quantified by means of a temporal first-order autocorrelation function (ACF):
	\begin{equation}\label{eq:acf1}
			g(\tau)=\frac{\mean{I(t)\cdot I(t+\tau)}_t}{\mean{I(t)}_t^2},\ \ \ \mean{I(t)}_t:=\lim\limits_{\tilde{T}\rightarrow\infty}\frac{1}{\tilde{T}}\int_0^{\tilde{T}}I(t)\;\mathrm{d}t
	\end{equation}
  The ACF usually contains features that are spread over several orders of magnitude in time (nanoseconds to seconds). 

   The standard FCS setup uses a confocal microscope in combination with single-photon sensitive detectors to acquire the fluorescence time trace $I(t)$ from one focal volume. Then the data are fed into a ``correlator'' (hardware or software component), which estimates the ACF over a certain dynamic range. 

 	 	 In the accompanying paper by \citeauthor{mocsar2012}\cite{mocsar2012}, a field programmable gate array (FPGA) implementation of such a correlator circuit for use with a confocal microscope and up to two single photon avalanche diode (SPAD) detectors is presented.  In recent years, the availability of fast cameras has triggered the development of different imaging modalities for FCS\cite{WOHLAND2007,DROSS2009a,WOHLAND2010}, which allow spatial mapping of the diffusion coefficient and other dynamic properties by calculating an ACF for each of potentially many pixels. This requires correlation hardware that can process the data stream very fast, ideally in real time, for all pixels of the image sensor. Here we extend the idea of hardware reuse, presented before\cite{mocsar2012}, for application to imaging FCS.

	 Our FPGA-based implementation can calculate 1024 autocorrelation functions in parallel and in real time.
	 The dynamic range of the ACFs is $\tau_{\text{min}}\ldots\tau_{\text{max}}=\unit[10]{\mu s}\ldots\unit[1]{s}$. As a data acquisition device we use the single photon avalanche diode array (SPAD array) \mbox{\textit{Radhard2}}
	 \cite{note1,carrara2009gamma}. From our sensor we read a $32\times32$ pixel frame every $\Delta t_{\text{frame}}=\unit[10]{\mu s}$ where each pixel contains $\unit[1]{bit}$ of information (no photon or at least one photon in the last $\Delta t_{\text{frame}}$). The design presented here could easily be adapted to other image sensors such as customized complementary metal oxide semiconductor (CMOS) or electron multiplying charge-coupled device (EMCCD) cameras.

	 \section{\mtau Hardware Correlators}
	 A hardware correlator estimates the ACF in \eqref{eq:acf1} from a finite sequence of intensity measurements 
	    \begin{equation}\label{eq:acf02} 
	    	I_n=\int_0^{\tau_{\text{min}}}I\left(n\cdot\tau_{\text{min}}+t\right)\mathrm{d}t,\ \ \ n=0,1,...T-1 
	    \end{equation}
	 with the number of samples $T$ and the integration time $\tau_{\text{min}}$ for one sample.
         When discretizing \eqref{eq:acf1} with this intensity sequence, care has to be taken not to bias the normalization $1/\mean{I}_t^{2}$. A viable choice is the ``symmetric normalization'' introduced in Ref.~\onlinecite{SCHAAETZEL1990}:
	 \begin{equation}\label{eq:acf2}
			\hat{g}_{\text{sym}}(\tau_k)
				=\frac{\frac{1}{T-\tau_k}\cdot\overbrace{\sum\limits_{n=\tau_k}^{T-1}I_n\cdot I_{n-\tau_k}}^{=:G_{\tau_k}}}{\biggl[\frac{1}{T}\cdot\underbrace{\sum\limits_{n=0}^{T-1}I_{n}}_{=:M_0}\biggr]\cdot\biggl[\frac{1}{T-\tau_k}\cdot\underbrace{\sum\limits_{n=\tau_k}^{T-1}I_{n-\tau_k}}_{=:M_{\tau_k}}\biggr]}
	 \end{equation}
         with a given set of lag times $\tau_k\in\mathbb{N}$ (in units of $\tau_{\text{min}}$, so $\tau=\tau_k\cdot\tau_{\text{min}}$). 
 	 When the full sequence $\{I_n\}_{n=0\ldots T-1}$ is available after the measurement, \eqref{eq:acf2} may be evaluated directly for an arbitrary (also logarithmically spaced) set of lags $\tau_k$. This gives an unbiased estimation of the ACF (``direct correlation''). 

\begin{figure*}[B]
\centering
\includegraphics[width=0.8\textwidth]{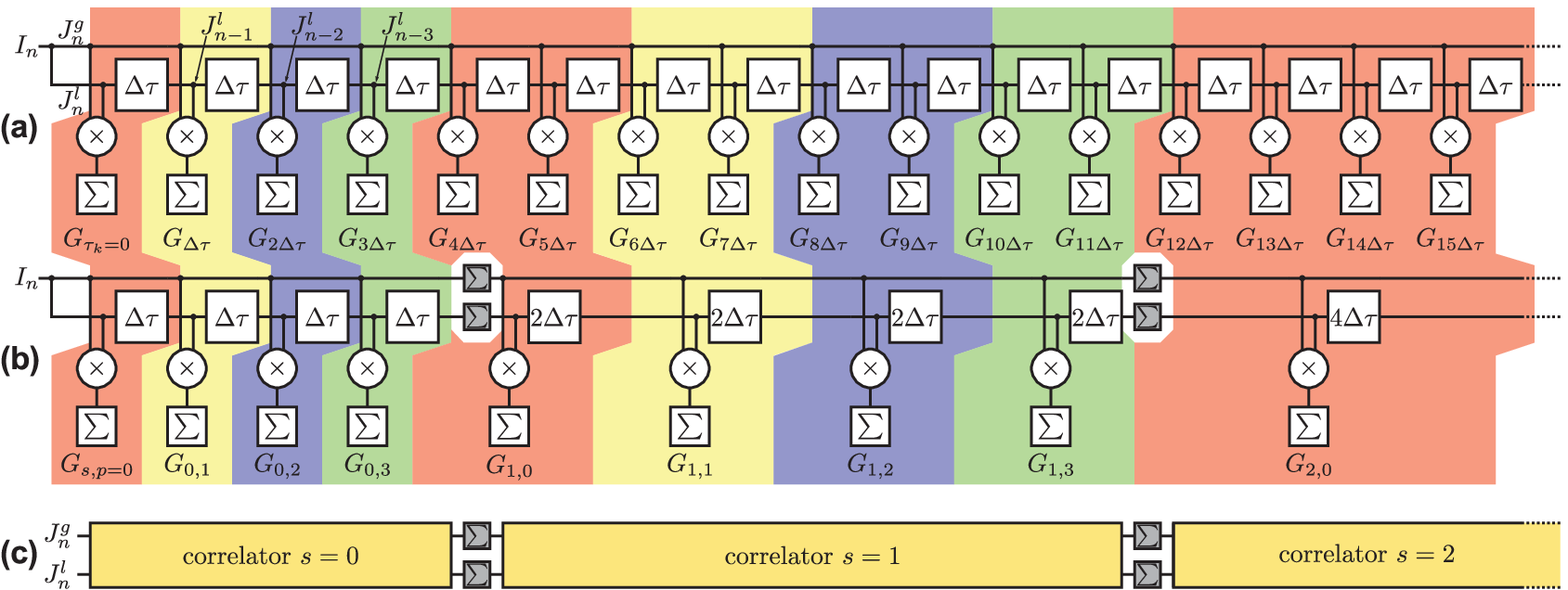}
\caption{Hardware design of a linear correlator (a) in comparison with a \multitau correlator (b). Panel (c) shows a schematic view of the \mtau correlator, where the linear correlator building blocks are summarized by a single block. Corresponding channels are grouped with the same color. Global/undelayed ($g$) and local/delayed ($l$) inputs are located on the left. For autocorrelation, the global and local signal inputs to the $0$-th block ($J_n^{(g)}=J_n^{(l)}=I_n$) are identical. The {\tiny $\boxed{\Delta\tau}$} blocks represent delay elements, the $\otimes$-blocks multipliers and the {\tiny $\boxed{\Sigma}$}-blocks accumulators.}
\label{fig:mtaucorr}
\end{figure*}

	To implement our hardware correlator, we use the \multitau scheme introduced in \lcite{SCHAAETZEL1985}, which is also illustrated and compared to a linear implementation in \Figref{fig:mtaucorr} (for a detailed description, please refer to our accompanying paper \lcite{mocsar2012}). The \mtau scheme uses a set of $S$ ``linear" correlator blocks (\figref{fig:mtaucorr}(b,c)). The input samples $I_{s,n}$ ($n$ is the same index as in \eqref{eq:acf02}) are  summed over increasingly long periods $\Delta n=m^s$: 
	 		\begin{equation}\label{eq:acf3}
	 		   I_{s,n}=\sum\limits_{k=1}^{m^s}I_{n-k},\ \ \ \ \ \text{for}\ \ s>0
	 		\end{equation}
	 with $I_{0,n}=I_n$.
	 
	 Each of the linear correlators estimates the ACF at $P$ linearly spaced lags 
\begin{align}
	\tau_{0,0} &= 0\notag\\
	\tau_{s,0} &= \tau_{s-1,P-1}+m^{s-1}\notag\\
	\tau_{s,p} &= \tau_{s,p-1}+m^s=\sum\limits_{i=1}^{s\cdot P+p}m^{\left\lfloor \frac{i-1}{P}\right\rfloor},\label{eq:acf4}
\end{align}
where $p=0\ldots P-1$.  

In summary, this results in a quasi-logarithmic spacing of estimates $\hat{g}_{\text{sym,\mtau}}(\tau_{s,p})$. The advantage of this \multitau scheme is its simple implementation in hardware and a large dynamic time range with a reasonable number of channels. Its disadvantage is a systematic error introduced by averaging: As shown in Ref.~\onlinecite{KOJRO1999} the estimator $\hat{g}_{\text{sym,\mtau}}(\tau_{s,p})$ equals the ideal correlation function $g(\tau_{s,p}\cdot\tau_{\text{min}})$ (see \eqref{eq:acf2}) convolved with a triangular kernel with width $m^s$:
	 		\begin{equation}\label{eq:acf3}
	 		   \hat{g}_{\text{sym,\mtau}}(\tau_{s,p})=g(\tau_{s,p}\cdot\tau_{\text{min}})\ast\Lambda(\tau_{s,p}, m^s),
\end{equation}
where $\ast$ denotes the convolution product and $ \Lambda(\tau, \Delta\tau)=\Delta\tau-\abs{\tau}$ for $\abs{\tau}< \Delta\tau$ and $\Lambda(\tau, \Delta\tau)=0$ for $\abs{\tau}\geq\Delta\tau$, is the triangular shaped kernel.

%
%
%
%
%

\section{Hardware design}
Here we describe how the hardware reuse scheme in the accompanying paper \lcite{mocsar2012} can be extended to accommodate many more input channels. Instead of the graphical tool used there, here we employed a low-level hardware description language to gain speed and flexibility. This enables us to fine-tune many parameters of the final design, e.g. operational speed, memory usage, logic resource consumption and routing between logic cells.

\subsection{Single-Pixel Correlator}
As shown in \figref{fig:mtaucorr}, a typical correlator is made up from channels, each corresponding to a certain lag time and consisting of a multiplier, an accumulator and a delay element.

The idea of our implementation is to use one single channel circuit to calculate all channels within one \mtau correlator. This is possible by serial processing of the lag time channels, since every channel's hardware is identical.

The basic arithmetic operation of one channel is to multiply and accumulate (MAC). Therefore we can map its functionality onto a MAC unit which can be found on most FPGA architectures and which is considerably faster than using generic FPGA logic cells. Only about 100 of these can be found on typical FPGAs, precluding any approach with blocks consisting of several lag channels.

As only one circuit is used to process all channels, we use an internal memory block (block random access memory, BRAM) to store their state (i.e., the content of the accumulator and the delayed signal). We implement this circuit by decomposing it into five steps:
\begin{enumerate}
		\item\textit{\textbf{L}oad} accumulator and delayed value of a channel from memory
		\item\textit{\textbf{W}ait} for memory access to complete
		\item\textit{\textbf{M}ultiply} delayed with global signal
		\item\textit{\textbf{A}dd} multiplication result to channel's accumulator
		\item\textit{\textbf{S}tore} counter and new delayed value to memory
\end{enumerate}
To increase performance, these steps are executed in an interleaved manner for four channels simultaneously. This ``pipelining'' scheme is shown in \tabref{tab:pipeline_int}. Our five pipeline steps are compatible with the internal pipelining of common MAC units.





\begin{table}[h]
	\caption{\label{tab:pipeline_int}Interleaved pipeline of the linear correlator design with 8 channels.}
	
	\begin{tabular}{|l|c@{ }c@{ }c@{ }c@{ }c@{ }c@{ }c@{ }c||c@{ }c@{ }c@{ }c@{ }c@{ }c@{ }c@{ }c|r|}
	\hline
 	cycle $c$ & 0 & 1 & 2 & 3 & 4 & 5 & 6 & 7 & 8 & 9 & 10 & 11 & 12 & 13 & 14 & 15 & \\ \hline \hline
	ch. 0 & L & W & M & A & S &   &   &   & L & W & M & A & S &   &   &   & ch. 4\\
	\hline
	ch. 1 &   & L & W & M & A & S &   &   &   & L & W & M & A & S &   &   &  ch. 5\\
	\hline
	ch. 2 &   &   & L & W & M & A & S &   &   &   & L & W & M & A & S &   & ch. 6 \\
	\hline
	ch. 3 &   &   &   & L & W & M & A & S &   &   &   & L & W & M & A & S & ch. 7\\
	\hline
	\end{tabular}
\end{table}

From one block to the next the delay time is doubled ($m=2$) in the \mtau scheme, making the input data rate of block $s+1$ half that of block $s$. Hence we need to execute each block only half as often as its predecessor. Thus, a complete \mtau correlator can be executed in only twice the run-time $\Delta t_{\text{lin}}$ of a single linear correlator block:
\begin{multline}\label{eq:correstimate1}
\overbrace{1\cdot\Delta t_{\text{lin}}}^{\text{1\textsuperscript{st} lin. corr.}}+
\overbrace{\frac{1}{2}\cdot\Delta t_{\text{lin}}}^{\text{2\textsuperscript{nd} lin. corr.}}+
\overbrace{\frac{1}{4}\cdot\Delta t_{\text{lin}}}^{\text{3\textsuperscript{rd} lin. corr.}}+\dots\\\le
\sum_{n=0}^{\infty} \frac{1}{2^n}\cdot\Delta t_{\text{lin}}=2\cdot\Delta t_{\text{lin}}
\end{multline}
A key requirement is that $\Delta t_{\text{lin}}$ is at most half the integration time $\tau_{\text{min}}$ of the input signal $I_n$. 



\begin{figure*}[b!]
\centering
\includegraphics[width=\textwidth]{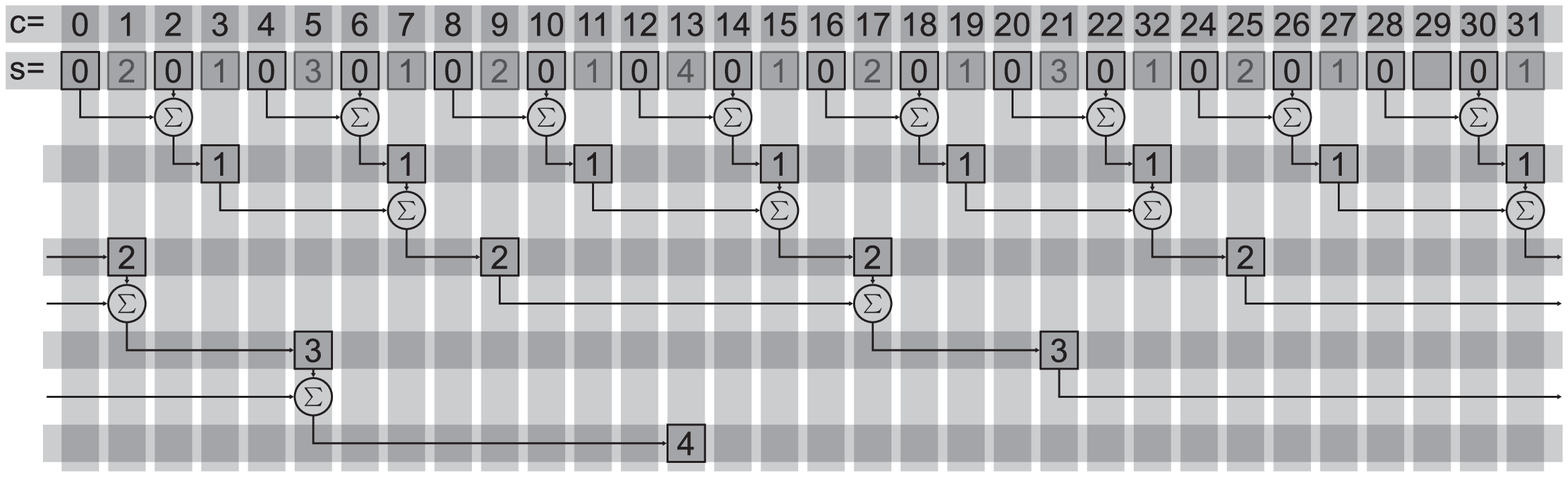}
\caption{Block scheduling algorithm. The counter $c$ in the first row defines the current cycle. The second row shows the block $s$ that is processed in the corresponding cycle. Further below the order of processing is shown: When a block $s$ has been processed twice, the following block, $s+1$, can process the sum of the two preceding $s$-blocks.}\label{fig:blocksched}
\end{figure*}

As before\cite{mocsar2012}, a scheduler guarantees that a linear correlator block $s$ is only executed after its predecessor $s-1$ has been executed twice. A counter $c=0,1,\ldots$ is incremented with every execution of any linear correlator block. The scheduler uses the following relations to determine which linear correlator block $s$ has to be executed at a given counter value $c$ (details see appendix):
\begin{align}
  s=0:   & & \text{c}\bmod2^{1}   &= 0\notag\\
  s=1:   & & \text{c}\bmod2^{2}   &= 3\label{eq:rel_counters}\\
  s\ge2: & & \text{c}\bmod2^{s+1} &= (2^s-3)\notag
\end{align}
\figref{fig:blocksched}(a) shows the solution of this relation for $c$ values from $0$ to $31$. In the binary representation of $c$ for a linear correlator block $s$ patterns are evident that can be used to implement the scheduler efficiently. As shown in \tabref{tab:bits} (for $S=8$ linear correlators), correlator block $s=0$ is executed whenever the last bit of $c$ is $0_{\text{b}}$, correlator $s=1$ is executed when the last two bits are $11_{\text{b}}$ and so forth. This scheme uses only simple comparison operations.
\begin{table}[t!]
	\caption{\label{tab:bits}
		Binary representation of counter $c$ that solves relations \eqref{eq:rel_counters} for a given linear correlator block $s$. \texttt{*} denotes a don't care condition. }
	\	
	\small
	\tabcolsep=0.1cm
	\begin{tabularx}{.79\linewidth}{|c||ccccccccccl|}
			\hline
			 &\multicolumn{11}{X|}{\centering binary representation of counter $c$ that solves relation \eqref{eq:rel_counters}}\\
		 lin.	corr. $s$& {\footnotesize $2^8$}      &&{\footnotesize$2^7$}     &{\footnotesize$2^6$}     &{\footnotesize$2^5$}     &{\footnotesize$2^4$}     &&{\footnotesize$2^3$}     &{\footnotesize$2^2$}     &{\footnotesize$2^1$}     &{\footnotesize$2^0$} \\ \hline\hline
			0 &\texttt{*}&&\texttt{*}&\texttt{*}&\texttt{*}&\texttt{*}&&\texttt{*}&\texttt{*}&\texttt{*}&\texttt{0} \\ 
			1 &\texttt{*}&&\texttt{*}&\texttt{*}&\texttt{*}&\texttt{*}&&\texttt{*}&\startt{0}&\texttt{1}&\texttt{1} \\ 
			2 &\texttt{*}&&\texttt{*}&\texttt{*}&\texttt{*}&\texttt{*}&&\startt{1}&\texttt{0}&\texttt{0}&\texttt{1} \\ 
			3 &\texttt{*}&&\texttt{*}&\texttt{*}&\texttt{*}&\startt{1}&&\texttt{0}&\texttt{1}&\texttt{0}&\texttt{1} \\ 
			4 &\texttt{*}&&\texttt{*}&\texttt{*}&\startt{1}&\texttt{0}&&\texttt{1}&\texttt{1}&\texttt{0}&\texttt{1} \\ 
			5 &\texttt{*}&&\texttt{*}&\startt{1}&\texttt{0}&\texttt{1}&&\texttt{1}&\texttt{1}&\texttt{0}&\texttt{1} \\ 
			6 &\texttt{*}&&\startt{1}&\texttt{0}&\texttt{1}&\texttt{1}&&\texttt{1}&\texttt{1}&\texttt{0}&\texttt{1} \\ 
			7 &\startt{1}&&\texttt{0}&\texttt{1}&\texttt{1}&\texttt{1}&&\texttt{1}&\texttt{1}&\texttt{0}&\texttt{1} \\ \hline
	\end{tabularx}
\end{table}

Between two consecutive blocks $s-1$ and $s$, adder circuitry is inserted to sum up two subsequent input signal values $I_{s-1,n-1}$ and $I_{s-1,n}$. This is done for both the delayed/local as well as the undelayed/global signal, while they are processed in the pipeline. 


The linear correlator as implemented here, together with its scheduler and the summation logic, is called correlation processing element (CorrPE).
All channel data and intermediate summation results, the so-called pixel context, are stored in a dual-port BRAM which is associated with the CorrPE.

\subsection{Multi-Pixel Correlator}
\begin{figure*}[h!]
	\centering
		\includegraphics[width=0.8\textwidth]{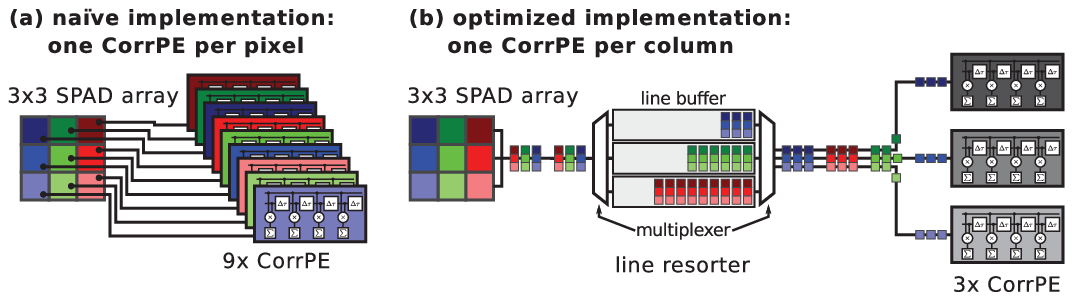}
	\caption{Comparison of a na\"ive implementation (a: one correlator per pixel) and an optimized implementation (b: reuse CorrPEs for several pixels) of a multi-pixel \mtau correlator for a $3\times3$ SPAD array.}
	\label{fig:corrpe}
\end{figure*}

The CorrPE described above is much faster than required to calculate the ACF for a single pixel in the SPAD array. Hence, we can reuse a single CorrPE for multiple pixels by switching between pixel contexts. This ``pixel scheduler'' uses a double-buffering strategy. While a \text{CorrPE} operates on the current context, the previous context is exchanged with the next context to be processed. The pixel contexts are stored in external background memory (SRAM), because they exceed the capacity of the internal memory. In \figref{fig:corrpe} we illustrate how we reuse one CorrPE for several pixels in comparison to a na\"ive implementation using one CorrPE per pixel.

In addition to the channel data, the cycle counter $c$ and an accumulator for the local and the global input signals have to be saved. The latter are used for normalization and cross-correlation.

\begin{figure*}[B]
\centering
\includegraphics{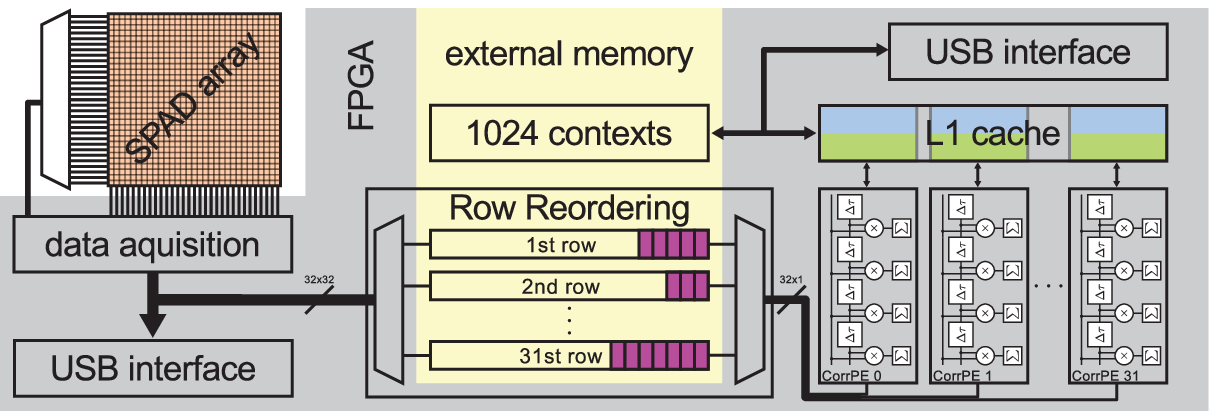}
\caption{System layout and data path - from data acquisition to correlation. One USB interface is used to stream the raw images, the other for streaming (intermediate) results. The first level cache (L1, double buffered) is used to hold the context of the currently processed pixel. FIFOs for row resorting and for context storage use external memory.}
\label{fig:system}
\end{figure*}

A single CorrPE is used to process an entire column (ACFs of $n_y=32$ pixels) of our SPAD array. To handle the full $n_x\times n_y$ array of pixels, we instantiate $n_x=32$ CorrPEs in parallel. An overview of this scheme is shown in \figref{fig:system}. A ``data acquisition'' circuit communicates with the SPAD array and provides the image data for the correlators. As the image data are streamed out row by row, and each of the $n_x$ CorrPEs is only processing data from one specific context (i.e. a specific row), the remaining pixels have to be buffered in 32 FIFOs (first-in first-out memory buffer) localized in external RAM.

In addition our design contains two USB 2.0 interfaces. One is used to send the raw data stream from the SPAD array to the computer, which allows further data processing. We also use these raw data to verify our correlator design. Via the second interface, intermediate and final results from the correlators are transferred to the host computer. The intermediate results allow implementing a live view of the ongoing calculation of the ACFs.



\subsection{ACF Normalization}
The intensity values in the denominator of \eqref{eq:acf2} are typically obtained from \textit{monitor channels} $M_{\tau_k}$, which accumulate the total photon count at a given lag time $\tau_k$.
In contrast to other implementations, we use only one monitor $M_0$ (input signal accumulator) per \multitau correlator. Symmetric normalization (see also \eqref{eq:acf2}) yields the following:
\begin{equation}\label{eq:norm1}
  \hat{g}_{\text{sym,\mtau}}(\tau_{s,p})= \frac{G_{\tau_{s,p}}}{2^s}\cdot\frac{T}{M_0\cdot M_{\tau_{s,p}}}
\end{equation}

After measuring $T$ samples, the number of samples that have propagated through the correlator to a distinct channel $(s,p)$ is $T-\tau_{s,p}$. Under the assumption that $I_n$ is a stationary random process and that $T>\tau_{s,p}$, the per-channel monitors $M_{\tau_{s,p}}$ can be estimated in the following manner:

\begin{equation}\label{eq:norm2}
  M_{\tau_{s,p}}=M_0\cdot\frac{T-\tau_{s,p}}{T}.
\end{equation}
This normalization and model fitting is done on the host computer, since floating-point arithmetic cannot be implemented efficiently on an FPGA. Although the correlation on the FPGA and the normalization on the host computer can easily be done in real time, the fits in most cases cannot. Usually a single curve fit takes tens of milliseconds, thus fitting all $1024$ ACFs would amount to $\ge\unit[10]{s}$ additional computing time; but  this still allows to display fit parameter maps (images) within an acceptably short delay. In current imaging FCS software systems\cite{sankaran2010imfcs} the data are loaded and correlated on the host computer, which for a measurement of typically $\unit[10]{s}$ takes $\geq\unit[10]{min}$ for $1024$ pixels at the frame rate of our sensor.

\begin{figure*}[b]
	\centering
		\includegraphics{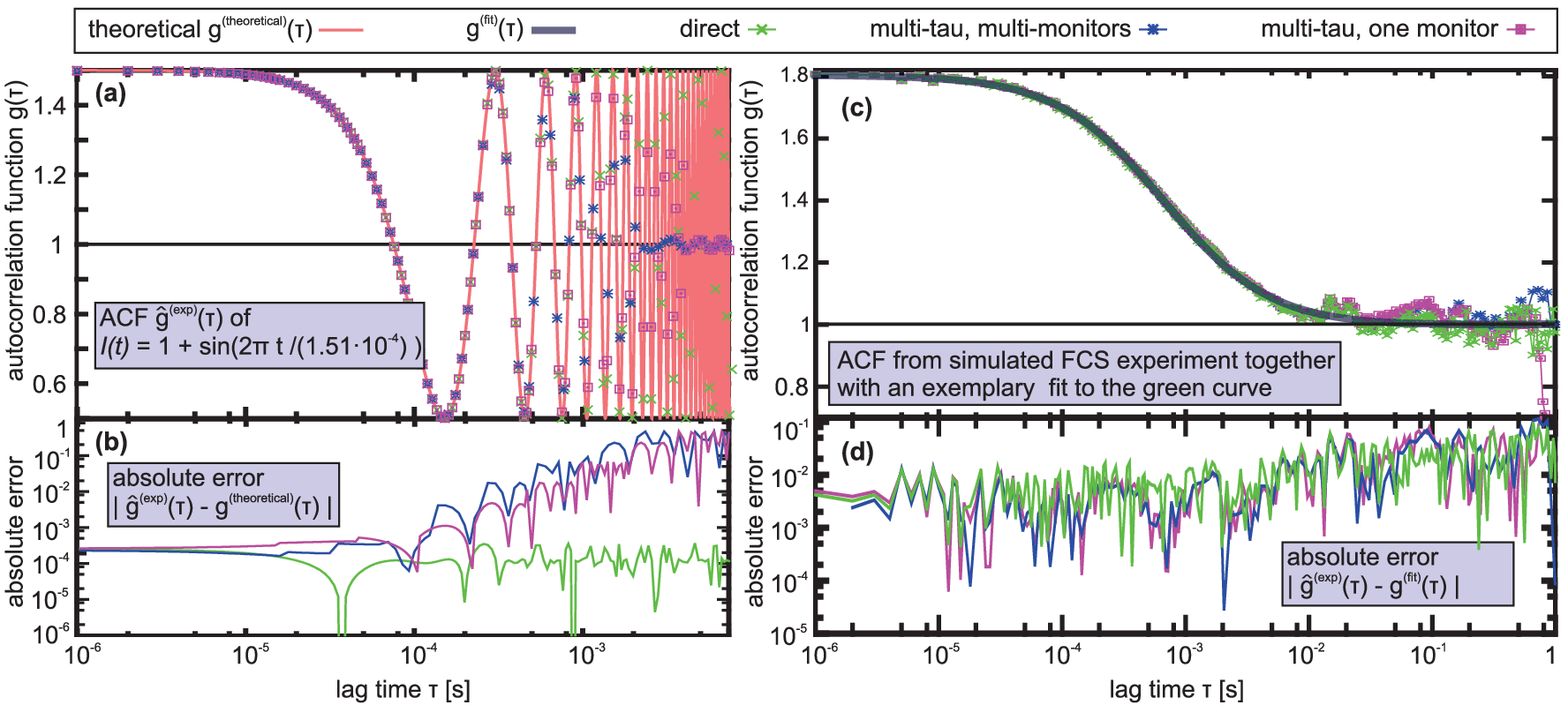}
	\caption{Simulation results for different implementations of \mtau correlators: The left panels (a,b) show simulations for a sine wave input signal $I(t)=1+\sin(2\pi t/(\unit[0.151]{ms}))$. The simulations on the right (c,d) were created using an FCS simulation. The top graphs (a,c) are estimates of the autocorrelation function using direct correlation from \eqref{eq:acf2} (green), a \mtau correlator with one monitor channel per lag time (blue) and our estimated normalization from \eqref{eq:norm2} (magenta). Graph (a) also shows the theorectical ACF $g^{\text{(theoretical)}}(\tau)=1+\cos(2\pi \tau/(\unit[0.151]{ms}))$ for the sine signal (light red). The lower graphs (b,d) show the absolute  deviation of the estimates from $g^{\text{(theoretical)}}(\tau)$ (b) and from a fit to the curves (d). The parameters resulting from the fit in (c) are the same within $<2.5\%$ for all three estimates. }
	\label{fig:corr_simulation}
\end{figure*}

To show that the estimation in \eqref{eq:norm2} yields good results, we simulated different correlator types in software. The results for a direct estimation of the ACF using \eqref{eq:acf2} (green), a \mtau correlator with a monitor channel per lag (blue) and our estimation (magenta) can be seen in \figref{fig:corr_simulation}, where the data in (a) and (b) were obtained by correlating the input signal $I(t)=1+\sin(2\pi t/(1.51\cdot 10^{-4}))$ for which the exact ACF is known to be $g^{\text{(theoretical)}}(\tau)=1+\cos(2\pi \tau/(1.51\cdot 10^{-4}))$ (time $t$ and lags $\tau$ are unit free). The data in \figref{fig:corr_simulation}(c) was created by simulating a $T_{\text{sim}}=\unit[1]{s}$ long FCS experiment with one diffusing species\cite{note2}.
It was computed with our FCS simulation software described in Ref.~\onlinecite{wocjan2009dynamics}. Further details on the simulation code are shown in the appendix. 

For short lags the estimated ACFs resemble the theoretical curves quite well. \Mtau correlators have an increased absolute error for longer lags, which is due to the averaging described in \eqref{eq:acf3}. This can  be seen especially in the case of the sine wave signal. The \mtau estimates can still be used for FCS experiments, as here the ACFs usually decay to 1 (white noise) for large lag times, and thus the systematic error drops to zero again (for a detailed discussion of this, see e.g. Ref.~\onlinecite{KOJRO1999}). For $\tau\gtrsim T_{\text{sim}}/10$, \mtau correlators show additional systematic deviations from the theoretical curve and from the direct estimation, because the channels are not averaged over sufficiently many samples to yield reliable results. Here the \mtau implementation with multiple monitor channels performs better due to the better estimation of the normalization factor $M_{\tau_{s,p}}$.

\subsection{Crosscorrelation (CCF)}
Our design can also calculate cross-correlation:
\[ g^{(x,y)}(\tau)=\frac{\mean{I^{(x)}(t)\cdot I^{(y)}(t+\tau)}_t}{\mean{I^{(x)}(t)}_t\cdot\mean{I^{(y)}(t)}_t} \]
between two input signals $I^{(x)}(t)$ and $I^{(y)}(t)$ at the local $J^{\text{(l)}}$ and global $J^{\text{(g)}}$ inputs of the CorrPE (see \figref{fig:mtaucorr} where both signals are tied to $I(t)$ for ACF calculation). This changes the \mtau estimator \eqref{eq:norm1} to:
\begin{equation}\label{eq:norm1ccf}
  \hat{g}^{\text{(CCF)}}_{\text{sym,\mtau}}(\tau_{s,p})= \frac{G_{\tau_{s,p}}}{2^s}\cdot\frac{T}{M^{(g)}_0\cdot M^{(l)}_{\tau_{s,p}}}
\end{equation}
Here the normalization uses the monitors $M^{(g)}_0$ for the global and $M^{(l)}_0$ for the local signal ($M^{(l)}_{\tau_{s,p}}=M^{(g)}_0\cdot(T-\tau_{s,p})/T$). For autocorrelation (see above) only one monitor channel is needed, as $M_0=M^{(g)}=M^{(l)}$.

%
%
%
%
%
%
%
%
%

\section{Performance \& Implementation Details}

The complete design is implemented in two \mbox{Virtex-II} Pro FPGAs (XC2VP40, Xilinx, \mbox{\url{http://www.xilinx.com/}}, San Jose, USA), on a LASP development board\cite{niclass2008}. One FPGA is used for data acquisition and line reordering, while the other one is used to implement the correlators. 
The total resource consumption within the second FPGA is around $80\%$. Using this hardware platform, there are currently $P=8$ channels within each of the $S=14$ linear correlator blocks. 

In \eqref{eq:correstimate1} we showed that the complete \mtau correlator can be executed within twice the time needed for the first linear correlator block, so we devote half of the execution time to the first linear correlator and the rest to the remaining blocks. Therefore a new input sample can be accepted only once every $2 \Delta t_\text{lin}$. Here $\Delta t_\text{lin} = \unit[2P+3]{cycles}$ is the time needed to process a new input sample $I_n$ in the first linear correlator block. The $3$ additional cycles are used for data hand over to the next block. 

The CorrPEs run with a clock frequency of $\unit[144]{MHz}$, which corresponds to an execution time of $\unit[264]{ns} = 2\Delta t_\text{lin}$. This is the minimum timespan between two subsequent samples, if no pixel multiplexing is used. Hence, our current FPGA platform can calculate $32$ different ACFs or CCFs with a minimum lag time of $\unit[264]{ns}$, which is comparable to the $\unit[100]{ns}$ designs presented in Ref.~\onlinecite{JAKOB2006} and more recently in Ref.~\onlinecite{mocsar2012}. Trading time resolution for more correlation functions, we can process all $1024$ pixels of the SPAD array at a frame rate of $\unit[100]{kfps}$ in real time.

To estimate the memory consumption of our design, we first look at the pixel context, which consists of $128$ words of $64$ bits each. 
\tabref{tab:MemoryLayoutOfAPixelContext} shows a detailed memory layout. Sixteen of the upper $\unit[32]{bits}$ of the raw accumulators store the current value of the delay registers $J^{(l)}_{s,p}$. The lower $\unit[32]{bits}$ contain the accumulator $G_{\tau_{s,p}}$. The monitor channels are $\unit[32]{bits}$ each. Data handover between consecutive blocks (accumulated local and global signals) is done via $\unit[16]{bit}$-wide memory areas, which is sufficient for $S\le16$. Since in the later linear correlators the accumulated input signals $I_{s,n}$ are multiplied, the sums $G_{\tau_{s,p}}$ and also the $I_{s,n}$ can get relatively large and may not fit in the $\unit[32]{bit}$ memory locations available. However, for our FCS application we estimate that the word sizes used are sufficient, since the per-pixel input event rate is limited. For a deeper discussion of this topic, see Ref.~\onlinecite{hoppe2001design}.

\begin{table}
	\centering
	\begin{tabular}{|r|p{65mm}|}
	  \hline
	  \textbf{data word} & \textbf{usage}\\\hline\hline
	  $0...111$ & delay register $J^{(l)}_{s,p}$ and raw accumulator values $G_{\tau_{s,p}}$\\
	  $112$ & status counter $c$ \\
	  $113...125$ & intermediate results for accumulated input signals $I_{s,n}$\\
	  $126$	& global monitor channel $M^{(g)}_0$\\
	  $127$ & local monitor channel $M^{(l)}_0$\\\hline
	\end{tabular}
	\caption{\label{tab:MemoryLayoutOfAPixelContext}Memory layout of a pixel context}
	
\end{table}

For our $32\times32$ pixel SPAD array the pixel contexts are stored in $32\cdot32\cdot128\cdot \unit[64]{bits}=\unit[512]{KBytes}$ of external SRAM. A second SRAM stores the $\unit[256]{KBytes}$ used for the FIFOs ($2048$ entries each) that hold the pixel data until they are processed.

The correlator design is implemented in VHDL (very high speed integrated circuit hardware description language). It can be configured via generics (a VHDL feature). Thus we can reuse our design for different needs (e.g. for different sensor sizes and frame rates). We expect our design to show a significant increase in performance when implemented on newer FPGA generations (e.g. Virtex 5 or Virtex 6 from Xilinx). 

To ensure the functional correctness of the designed correlator, a simulation of the hardware was tested using random input data. The outcome was compared to the results of a software implementation of the \mtau correlator using the same data set. Both yielded exactly the same results.
The comparison was also done using real experimental data from the SPAD array. Again, both results were identical. 

\begin{figure}
	\centering
		\includegraphics{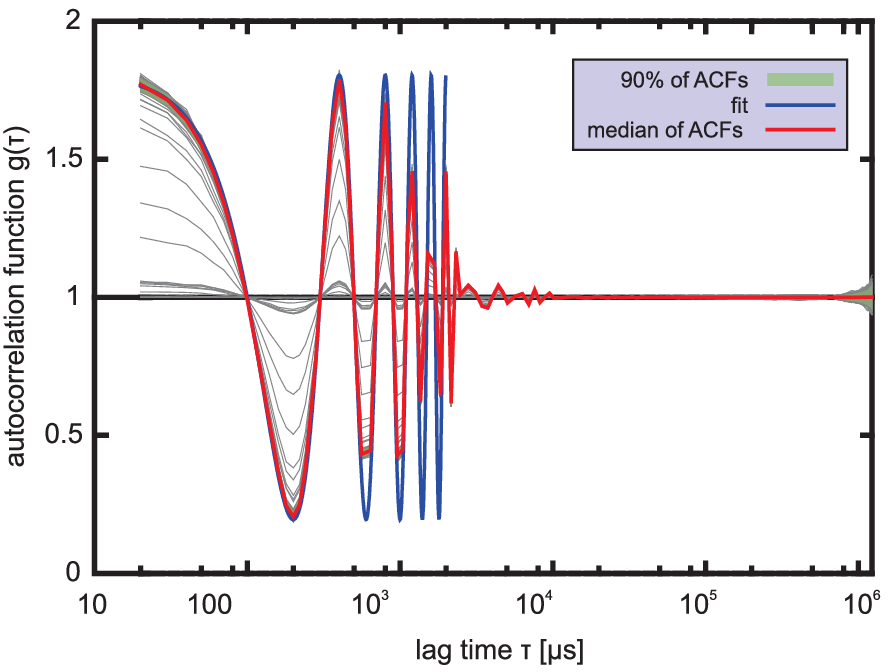}
	\caption{Distribution of 992 (gray) ACFs taken by our
sensor (first 31 columns only), exposed to a 630 nm LED
sine-modulated with a frequency of $\unit[2.5]{kHz}$. The correlator was running for $\unit[1.2]{s}$ (131072 samples at $\tau_\text{min}=\unit[10]{\mu s}$). The gray curves with significantly lower amplitude are due to hot pixels of the SPAD array; since these SPADs
fire randomly, the corresponding correlation amplitude is decreased. The median, which tends to be less sensitive towards outliers than the mean, is shown in red. The
theoretical model $g^{\text{(theoretical)}}(\tau)=1+A\cdot\cos(2\pi f\cdot\tau)$ is fitted against the median in the interval $\left[\unit[10]{\mu s}, \unit[1]{ms}\right]$, and is shown until $\tau=\unit[2]{ms}$. The fit yields a frequency of $\unit[(2502\pm4)]{Hz}$. The fanning out of the curve at high values of $\tau$ is due to unfilled correlator channels. Compare also \figref{fig:corr_simulation}(a).}
	\label{fig:corr_expresults}
\end{figure}

To demonstrate the functionality of the whole system, we tested the design using the SPAD array to record an LED connected to a sine wave generator set to $\unit[2.5]{kHz}$. \figref{fig:corr_expresults} shows the results of this measurement. A fit to the data recovered the chosen  frequency. 

\section{Conclusion}
In this paper we presented the implementation of an FPGA-based \mtau correlator design that can calculate $1024$ correlation functions in real time at a minimum lag time of $\unit[10]{\mu s}$. To our knowledge this is the largest number of real-time \mtau correlators implemented so far in a single device. The minimum lag time of $\unit[10]{\mu s}$ in our design is considerably longer than that of currently available hardware correlators (e.g. from ALV GmbH, Langen, Germany or correlator.com, Bridgewater, USA and \lcite{MAGATTI2003}). However, those are limited to at most $32$ auto-correlators. 

We use our design to correlate the output of a single-photon avalanche diode array used as image sensor. This combination will allow us to perform imaging fluorescence correlation spectroscopy at sufficiently fast time scales to resolve even the motion of small molecules in solution and living cells.

Our design is flexible, so beside estimating ACFs, temporal cross-correlation functions of different pixels\cite{DERTINGE2007,Colyer2010} or multiple colors\cite{Bestvater2010} (using spectrally resolved detectors) can also be calculated.

\begin{acknowledgments}
The project was supported by a NUS-BW (National University of Singapore / Baden-W\"urttemberg) joint grant to J.L., 
a doctoral fellowship of the Helmholtz International Graduate School for Cancer Research to J.B., 
a doctoral fellowship of the Heidelberg Graduate School of Mathematical and Computational Methods for the Sciences to J.W.K.
We thank Xilinx, San Jose, USA for donating the FPGAs on the LASP development board.
G.V. receives support by the German-Hungarian program for the exchange of researchers by the German Academic Exchange Service and the Hungarian Scholarship Board (M\"OB-47-1/2010) OTKA K77600 and TAMOP 4.2.1/B-09/1/KONV-2010-007.
\end{acknowledgments}

\appendix
\section{Relation \eqref{eq:rel_counters}}

By definition, Eq.~\ref{eq:rel_counters} can be rewritten in the following manner, with $C_s$ containing the cycles in which the linear correlator $s$ is processed:
\begin{eqnarray*}
C_0 &=& \left\{c=n\cdot2^1+0           \mid c,n\in\mathbb{N}_0 \right\}\\
C_1 &=& \left\{c=n\cdot2^2+3           \mid c,n\in\mathbb{N}_0 \right\}\\
C_s &=& \left\{c=n\cdot2^{s+1}+2^{s}-3 \mid c,n\in\mathbb{N}_0 \right\} \forall s\ge2
\end{eqnarray*}

It remains to show that $C_p\cap C_q\forall p,q\in\mathbb{N}_0$, $p\neq q$. Obviously $C_0\cap C_s\;\forall s\in \mathbb{N}$,
$k,l \in\mathbb{N}_0$:
\begin{align*}
&&                C_1 \cap  C_2                      & = \varnothing\\
&\Rightarrow&     k\cdot2^2+3                        &\neq  l\cdot2^3+1\\
&\Leftrightarrow& \underbrace{k}_{\in\mathbb{N}_0} &\neq  \underbrace{2l}_{\in\mathbb{N}_0}-\underbrace{\frac{1}{2}}_{\in\mathbb{Q}}\\
&\Rightarrow&                                        &\surd
\end{align*}

Without limitations let $r<s$; $\forall r,s\in\mathbb{N}_0$; $r,s\ge2$; $k,l \in\mathbb{N}_0$:
\begin{align*}
&&                C_r \cap  C_s                      & = \varnothing\\
&\Rightarrow&     k\cdot2^{r+1}+2^{r}-3              &\neq  l\cdot2^{s+1}+2^{s}-3\\
&\Leftrightarrow& \underbrace{k}_{\in\mathbb{N}_0} &\neq  \underbrace{l\cdot2^{s-r} + 2^{s-r-1}}_{\in\mathbb{N}_0}-\underbrace{\frac{1}{2}}_{\in\mathbb{Q}}\\
&\Rightarrow&                                        &\surd
\end{align*}

\section{FCS Simulation \& Fits}
To test different types of autocorrelators, we used our FCS simulation program described in Ref.~\onlinecite{wocjan2009dynamics}. It simulates the 3D trajectories \[\vec{r}_i(t), t=0...T_{\text{sim}}, i=1..K\] of a set of $K$ fluorescing particles performing a random walk in which each 1D step is drawn from a zero-centered Gaussian distribution with width \[ \sigma_{\text{jump}}=\sqrt{2D\cdot\Delta t_{\text{sim}}}. \] Here $D$ is the given diffusion coefficient and $\Delta t_{\text{sim}}$ is the simulation time step. We use a Gaussian approximation for the focal illumination and detection volume:
\[
  h(x,y,z)=\exp\left(-2\cdot\frac{x^2+y^2}{w_{\rm xy}^2}-2\cdot\frac{z^2}{\gamma^2w_{\rm xy}^2}\right),
\]
where $w_{\rm xy}$ is the lateral width of the profile, $\gamma=w_{\rm z}/w_{\rm xy}$ is its aspect ratio and $w_{\rm z}$ is its length. The program then estimates the expected number of photons $\overline{N}_{\rm phot}(t)$ detected during each simulation time step $[t,t+\Delta t_{\rm sim}]$:
\[
  \overline{N}_{\rm phot}(t)=\overline{N}_0\cdot\Delta t_{\rm sim}\cdot\sum\limits_{i=1}^K q_{\rm f}\cdot q_{\rm det}\cdot h(\vec{r}_i(t))^2.
\]
Here $\overline{N}_0$ is the maximum number of detected photons per fluorophore and time step, while $q_{\rm f}$ and $q_{\rm det}$ are the quantum efficiencies of fluorescence and detection. To account for the counting statistics in the SPADs, the number of photons $N_{\rm phot}(t)$ actually detected during a simulation time step is calculated by drawing a random number from a Poissonian distribution with average $\overline{N}_{\rm phot}(t)$. 

The time series created by the simulation is then fed into the three different autocorrelator software implementations: (a) direct estimation, (b) \mtau with one monitor channel and (c) \mtau with multiple monitor channels. The latter two (b, c) simulate the complete structure of a \mtau correlator. In correlator (b) we use one single monitor channel, which counts the incoming photons only and then uses \eqref{eq:norm2} for the normalization. In the correlator (c) we use one monitor channel per correlator lag which counts the photons actually processed by each lag. 

\begin{figure}
	\centering
		\includegraphics{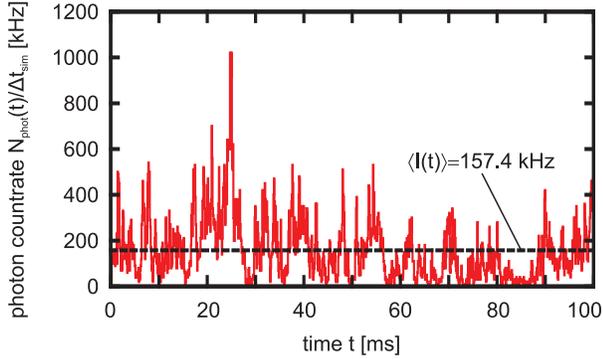}
	\caption{Typical photon count time trace $N_{\rm phot}(t)/\Delta t_{\rm sim}$ generated by our simulation}
	\label{fig:corr_simintensity}
\end{figure}

The simulation parameters for the data in \figref{fig:corr_simulation} are summarized in \tabref{tab:SimParameters}. The average photon count rate in the simulations was $\langle I(t)\rangle\approx \unit[157.4]{kHz}$ which is -- due to the chosen detection efficiency $q_{\rm det}=0.5$ -- about a factor of 5 higher than what would be expected from a standard confocal FCS setup. This way the signal to noise ratio is high enough to visualize easily artifacts introduced by the correlator implementation. \figref{fig:corr_simintensity} shows an example of a time trace $N_{\rm phot}(t)/\Delta t_{\rm sim}$ generated by our simulation.

\begin{table}
	\begin{tabular}{|p{10mm}|p{52mm}|p{18mm}|}
	 \hline
	 $ T_{\rm sim} $                  & duration of simulation               & $ 1\;\mathrm{s} $\\
	 $ \Delta t_{\rm sim} $           & simulation time step                 & $ 1\;\mathrm{\mu s} $\\
	 $ K $                            & number of walkers                    & $ 1576 $\\
	 $ V_{\rm sim} $                  & simulation volume                    & $ \unit[524]{\mu m^3} $\\
	 $ c $                            & concentration of simulated particles & $\unit[5]{nM}$  \\
	 $ D $                            & diffusion coefficient                & $\unit[20]{\mu m^2/s}$  \\
	 $ \overline{N}_0 $               & absorbed photons per molecule        & $4.2\cdot\unit[10^6]{s^{-1}}$ \\
	 $ w_{\rm xy} $                   & lateral width of excitation profile  & $ \unit[0.325]{\mu m} $  \\
	 $ q_{\rm f} $                    & fluorescence quantum yield           & $0.8$  \\
	 $ q_{\rm det} $                  & detection efficiency                 & $0.5$  \\
	 $ \gamma $                       & focal aspect ration                  & $6$ \\
	 $ S $                            & number of linear correlator blocks   & $20$ \\
	 $ P $                            & number of channels per block         & $16$ \\
	 $ m $                            & lag time ratio between blocks        & $2$ \\
	 \hline
  \end{tabular}
  
  \caption{\label{tab:SimParameters}Simulation parameters for the FCS simulation yielding the results, displayed in \figref{fig:corr_simulation}.}
\end{table}

The resulting ACFs were fitted to a standard 3D FCS model function\cite{KRICHEVS2002}:
\[ g^{\text{(fit)}}(\tau)=\frac{1}{N}\cdot\left(1+\frac{\tau}{\tau_D}\right)^{-1}\cdot\left(1+\frac{\tau}{\gamma^2\tau_D}\right)^{-1/2} \]
with the average particle number $N$ in the effective focal volume $V_{\rm eff}=\pi^{3/2}\cdot\gamma\cdot w_{\rm xy}^3$ and the diffusion decay time $\tau_D={w^2_{\rm xy}}/{4D}$.
The fits were performed using a Levenberg-Marquardt least squares fitting routine.

\section*{References \& Footnotes}


\begin{thebibliography}{22}%
\makeatletter
\providecommand \@ifxundefined [1]{%
 \@ifx{#1\undefined}
}%
\providecommand \@ifnum [1]{%
 \ifnum #1\expandafter \@firstoftwo
 \else \expandafter \@secondoftwo
 \fi
}%
\providecommand \@ifx [1]{%
 \ifx #1\expandafter \@firstoftwo
 \else \expandafter \@secondoftwo
 \fi
}%
\providecommand \natexlab [1]{#1}%
\providecommand \enquote  [1]{``#1''}%
\providecommand \bibnamefont  [1]{#1}%
\providecommand \bibfnamefont [1]{#1}%
\providecommand \citenamefont [1]{#1}%
\providecommand \href@noop [0]{\@secondoftwo}%
\providecommand \href [0]{\begingroup \@sanitize@url \@href}%
\providecommand \@href[1]{\@@startlink{#1}\@@href}%
\providecommand \@@href[1]{\endgroup#1\@@endlink}%
\providecommand \@sanitize@url [0]{\catcode `\\12\catcode `\$12\catcode
  `\&12\catcode `\#12\catcode `\^12\catcode `\_12\catcode `\%12\relax}%
\providecommand \@@startlink[1]{}%
\providecommand \@@endlink[0]{}%
\providecommand \url  [0]{\begingroup\@sanitize@url \@url }%
\providecommand \@url [1]{\endgroup\@href {#1}{\urlprefix }}%
\providecommand \urlprefix  [0]{URL }%
\providecommand \Eprint [0]{\href }%
\providecommand \doibase [0]{http://dx.doi.org/}%
\providecommand \selectlanguage [0]{\@gobble}%
\providecommand \bibinfo  [0]{\@secondoftwo}%
\providecommand \bibfield  [0]{\@secondoftwo}%
\providecommand \translation [1]{[#1]}%
\providecommand \BibitemOpen [0]{}%
\providecommand \bibitemStop [0]{}%
\providecommand \bibitemNoStop [0]{.\EOS\space}%
\providecommand \EOS [0]{\spacefactor3000\relax}%
\providecommand \BibitemShut  [1]{\csname bibitem#1\endcsname}%
\let\auto@bib@innerbib\@empty
\bibitem [{\citenamefont {Magde}, \citenamefont {Elson},\ and\ \citenamefont
  {Webb}(1974{\natexlab{a}})}]{MAGDE1974}%
  \BibitemOpen
  \bibfield  {author} {\bibinfo {author} {\bibfnamefont {D.}~\bibnamefont
  {Magde}}, \bibinfo {author} {\bibfnamefont {E.~L.}\ \bibnamefont {Elson}}, \
  and\ \bibinfo {author} {\bibfnamefont {W.~W.}\ \bibnamefont {Webb}},\ }\href
  {\doibase 10.1002/bip.1974.360130102} {\bibfield  {journal} {\bibinfo
  {journal} {Biopolymers}\ }\textbf {\bibinfo {volume} {13}},\ \bibinfo {pages}
  {1} (\bibinfo {year} {1974}{\natexlab{a}})}\BibitemShut {NoStop}%
\bibitem [{\citenamefont {Magde}, \citenamefont {Elson},\ and\ \citenamefont
  {Webb}(1974{\natexlab{b}})}]{MAGDE1974a}%
  \BibitemOpen
  \bibfield  {author} {\bibinfo {author} {\bibfnamefont {D.}~\bibnamefont
  {Magde}}, \bibinfo {author} {\bibfnamefont {E.~L.}\ \bibnamefont {Elson}}, \
  and\ \bibinfo {author} {\bibfnamefont {W.~W.}\ \bibnamefont {Webb}},\ }\href
  {\doibase 10.1002/bip.1974.360130103} {\bibfield  {journal} {\bibinfo
  {journal} {Biopolymers}\ }\textbf {\bibinfo {volume} {13}},\ \bibinfo {pages}
  {29} (\bibinfo {year} {1974}{\natexlab{b}})}\BibitemShut {NoStop}%
\bibitem [{\citenamefont {Krichevsky}\ and\ \citenamefont
  {Bonnet}(2002)}]{KRICHEVS2002}%
  \BibitemOpen
  \bibfield  {author} {\bibinfo {author} {\bibfnamefont {O.}~\bibnamefont
  {Krichevsky}}\ and\ \bibinfo {author} {\bibfnamefont {G.}~\bibnamefont
  {Bonnet}},\ }\href {\doibase 10.1088/0034-4885/65/2/203} {\bibfield
  {journal} {\bibinfo  {journal} {Reports on Progress in Physics}\ }\textbf
  {\bibinfo {volume} {65}},\ \bibinfo {pages} {251} (\bibinfo {year}
  {2002})}\BibitemShut {NoStop}%
\bibitem [{\citenamefont {Mocs\'{a}r}\ \emph {et~al.}(2011)\citenamefont
  {Mocs\'{a}r}, \citenamefont {Kreith}, \citenamefont {Buchholz}, \citenamefont
  {Krieger}, \citenamefont {Langowski},\ and\ \citenamefont
  {V\'{a}mosi}}]{mocsar2012}%
  \BibitemOpen
  \bibfield  {author} {\bibinfo {author} {\bibfnamefont {G.}~\bibnamefont
  {Mocs\'{a}r}}, \bibinfo {author} {\bibfnamefont {B.}~\bibnamefont {Kreith}},
  \bibinfo {author} {\bibfnamefont {J.}~\bibnamefont {Buchholz}}, \bibinfo
  {author} {\bibfnamefont {J.~W.}\ \bibnamefont {Krieger}}, \bibinfo {author}
  {\bibfnamefont {J.}~\bibnamefont {Langowski}}, \ and\ \bibinfo {author}
  {\bibfnamefont {G.}~\bibnamefont {V\'{a}mosi}},\ }\href@noop {} {\bibfield
  {journal} {\bibinfo  {journal} {Review of Scientific Instruments}\ }\textbf
  {\bibinfo {volume} {?}},\ \bibinfo {pages} {?} (\bibinfo {year}
  {2011})}\BibitemShut {NoStop}%
\bibitem [{\citenamefont {Kannan}\ \emph {et~al.}(2007)\citenamefont {Kannan},
  \citenamefont {Guo}, \citenamefont {Sudhaharan}, \citenamefont {Ahmed},
  \citenamefont {Maruyama},\ and\ \citenamefont {Wohland}}]{WOHLAND2007}%
  \BibitemOpen
  \bibfield  {author} {\bibinfo {author} {\bibfnamefont {B.}~\bibnamefont
  {Kannan}}, \bibinfo {author} {\bibfnamefont {L.}~\bibnamefont {Guo}},
  \bibinfo {author} {\bibfnamefont {T.}~\bibnamefont {Sudhaharan}}, \bibinfo
  {author} {\bibfnamefont {S.}~\bibnamefont {Ahmed}}, \bibinfo {author}
  {\bibfnamefont {I.}~\bibnamefont {Maruyama}}, \ and\ \bibinfo {author}
  {\bibfnamefont {T.}~\bibnamefont {Wohland}},\ }\href {\doibase
  10.1021/ac0624546} {\bibfield  {journal} {\bibinfo  {journal} {Analytical
  Chemistry}\ }\textbf {\bibinfo {volume} {79}},\ \bibinfo {pages} {4463}
  (\bibinfo {year} {2007})}\BibitemShut {NoStop}%
\bibitem [{\citenamefont {Dross}\ \emph {et~al.}(2009)\citenamefont {Dross},
  \citenamefont {Spriet}, \citenamefont {Zwerger}, \citenamefont {M\"{u}ller},
  \citenamefont {Waldeck},\ and\ \citenamefont {Langowski}}]{DROSS2009a}%
  \BibitemOpen
  \bibfield  {author} {\bibinfo {author} {\bibfnamefont {N.}~\bibnamefont
  {Dross}}, \bibinfo {author} {\bibfnamefont {C.}~\bibnamefont {Spriet}},
  \bibinfo {author} {\bibfnamefont {M.}~\bibnamefont {Zwerger}}, \bibinfo
  {author} {\bibfnamefont {G.}~\bibnamefont {M\"{u}ller}}, \bibinfo {author}
  {\bibfnamefont {W.}~\bibnamefont {Waldeck}}, \ and\ \bibinfo {author}
  {\bibfnamefont {J.}~\bibnamefont {Langowski}},\ }\href {\doibase
  10.1371/journal.pone.0005041} {\bibfield  {journal} {\bibinfo  {journal}
  {PLoS ONE}\ }\textbf {\bibinfo {volume} {4}},\ \bibinfo {pages} {e5041}
  (\bibinfo {year} {2009})}\BibitemShut {NoStop}%
\bibitem [{\citenamefont {Wohland}\ \emph {et~al.}(2010)\citenamefont
  {Wohland}, \citenamefont {Shi}, \citenamefont {Sankaran},\ and\ \citenamefont
  {Stelzer}}]{WOHLAND2010}%
  \BibitemOpen
  \bibfield  {author} {\bibinfo {author} {\bibfnamefont {T.}~\bibnamefont
  {Wohland}}, \bibinfo {author} {\bibfnamefont {X.}~\bibnamefont {Shi}},
  \bibinfo {author} {\bibfnamefont {J.}~\bibnamefont {Sankaran}}, \ and\
  \bibinfo {author} {\bibfnamefont {E.~H.~K.}\ \bibnamefont {Stelzer}},\ }\href
  {\doibase 10.1364/OE.18.010627} {\bibfield  {journal} {\bibinfo  {journal}
  {Optics Express}\ }\textbf {\bibinfo {volume} {10}},\ \bibinfo {pages}
  {10627} (\bibinfo {year} {2010})}\BibitemShut {NoStop}%
\bibitem [{not({\natexlab{a}})}]{note1}%
  \BibitemOpen
  \bibinfo {note} {The single SPADs of
  Radhard2 have a photon detection probability of around $35\%$ at a wavelength
  of $\unit[525]{nm}$. The fill factor of the array is around
  $1.5\%$.}\BibitemShut {Stop}%
\bibitem [{\citenamefont {Carrara}\ \emph {et~al.}(2009)\citenamefont
  {Carrara}, \citenamefont {Niclass}, \citenamefont {Scheidegger},
  \citenamefont {Shea},\ and\ \citenamefont {Charbon}}]{carrara2009gamma}%
  \BibitemOpen
  \bibfield  {author} {\bibinfo {author} {\bibfnamefont {L.}~\bibnamefont
  {Carrara}}, \bibinfo {author} {\bibfnamefont {C.}~\bibnamefont {Niclass}},
  \bibinfo {author} {\bibfnamefont {N.}~\bibnamefont {Scheidegger}}, \bibinfo
  {author} {\bibfnamefont {H.}~\bibnamefont {Shea}}, \ and\ \bibinfo {author}
  {\bibfnamefont {E.}~\bibnamefont {Charbon}},\ }in\ \href@noop {} {\emph
  {\bibinfo {booktitle} {ISSCC, IEEE International Solid-State Circuits
  Conference}}}\ (\bibinfo {year} {2009})\ pp.\ \bibinfo {pages}
  {40--41}\BibitemShut {NoStop}%
\bibitem [{\citenamefont {Sch\"{a}tzel}(1990)}]{SCHAAETZEL1990}%
  \BibitemOpen
  \bibfield  {author} {\bibinfo {author} {\bibfnamefont {K.}~\bibnamefont
  {Sch\"{a}tzel}},\ }\href {\doibase 10.1088/0954-8998/2/4/002} {\bibfield
  {journal} {\bibinfo  {journal} {Quantum Opt.}\ }\textbf {\bibinfo {volume}
  {2}},\ \bibinfo {pages} {287} (\bibinfo {year} {1990})}\BibitemShut {NoStop}%
\bibitem [{\citenamefont {Sch\"{a}tzel}(1985)}]{SCHAAETZEL1985}%
  \BibitemOpen
  \bibfield  {author} {\bibinfo {author} {\bibfnamefont {K.}~\bibnamefont
  {Sch\"{a}tzel}},\ }\href@noop {} {\bibfield  {journal} {\bibinfo  {journal}
  {Institute of Physics Conference Series}\ }\textbf {\bibinfo {volume} {77}},\
  \bibinfo {pages} {175} (\bibinfo {year} {1985})}\BibitemShut {NoStop}%
\bibitem [{\citenamefont {Kojro}\ \emph {et~al.}(1999)\citenamefont {Kojro},
  \citenamefont {Riede}, \citenamefont {Schubert},\ and\ \citenamefont
  {Grill}}]{KOJRO1999}%
  \BibitemOpen
  \bibfield  {author} {\bibinfo {author} {\bibfnamefont {Z.}~\bibnamefont
  {Kojro}}, \bibinfo {author} {\bibfnamefont {A.}~\bibnamefont {Riede}},
  \bibinfo {author} {\bibfnamefont {M.}~\bibnamefont {Schubert}}, \ and\
  \bibinfo {author} {\bibfnamefont {W.}~\bibnamefont {Grill}},\ }\href
  {\doibase 10.1063/1.1150101} {\bibfield  {journal} {\bibinfo  {journal}
  {Review of Scientific Instruments}\ }\textbf {\bibinfo {volume} {70}},\
  \bibinfo {pages} {4487} (\bibinfo {year} {1999})}\BibitemShut {NoStop}%
\bibitem [{\citenamefont {Sankaran}\ \emph {et~al.}(2010)\citenamefont
  {Sankaran}, \citenamefont {Shi}, \citenamefont {Ho}, \citenamefont
  {Stelzer},\ and\ \citenamefont {Wohland}}]{sankaran2010imfcs}%
  \BibitemOpen
  \bibfield  {author} {\bibinfo {author} {\bibfnamefont {J.}~\bibnamefont
  {Sankaran}}, \bibinfo {author} {\bibfnamefont {X.}~\bibnamefont {Shi}},
  \bibinfo {author} {\bibfnamefont {L.}~\bibnamefont {Ho}}, \bibinfo {author}
  {\bibfnamefont {E.}~\bibnamefont {Stelzer}}, \ and\ \bibinfo {author}
  {\bibfnamefont {T.}~\bibnamefont {Wohland}},\ }\href {\doibase
  10.1364/OE.18.025468} {\bibfield  {journal} {\bibinfo  {journal} {Optics
  Express}\ }\textbf {\bibinfo {volume} {18}},\ \bibinfo {pages} {25468}
  (\bibinfo {year} {2010})}\BibitemShut {NoStop}%
\bibitem [{not({\natexlab{b}})}]{note2}%
  \BibitemOpen
  \bibinfo {note} {The diffusion
  coefficient was $D=\unit[20]{\mu m^2/s}$ (corresponding to an intermediately
  sized protein in water), the simulation timestep of the random walk, as well
  as the minimum lag time were $\Delta
  t_{\text{sim}}=\tau_{\text{min}}=\unit[1]{\mu s}$. There were around $1.2$
  particles in the effective measurement volume
  $V_{\text{eff}}\approx\unit[0.4]{\mu m^3}$ on average.}\BibitemShut {Stop}%
\bibitem [{\citenamefont {Wocjan}\ \emph {et~al.}(2009)\citenamefont {Wocjan},
  \citenamefont {Krieger}, \citenamefont {Krichevsky},\ and\ \citenamefont
  {Langowski}}]{wocjan2009dynamics}%
  \BibitemOpen
  \bibfield  {author} {\bibinfo {author} {\bibfnamefont {T.}~\bibnamefont
  {Wocjan}}, \bibinfo {author} {\bibfnamefont {J.}~\bibnamefont {Krieger}},
  \bibinfo {author} {\bibfnamefont {O.}~\bibnamefont {Krichevsky}}, \ and\
  \bibinfo {author} {\bibfnamefont {J.}~\bibnamefont {Langowski}},\ }\href
  {\doibase 10.1039/B911857H} {\bibfield  {journal} {\bibinfo  {journal} {Phys.
  Chem. Chem. Phys.}\ }\textbf {\bibinfo {volume} {11}},\ \bibinfo {pages}
  {10671} (\bibinfo {year} {2009})}\BibitemShut {NoStop}%
\bibitem [{\citenamefont {Niclass}\ \emph {et~al.}(2008)\citenamefont
  {Niclass}, \citenamefont {Favi}, \citenamefont {Kluter}, \citenamefont
  {Gersbach},\ and\ \citenamefont {Charbon}}]{niclass2008}%
  \BibitemOpen
  \bibfield  {author} {\bibinfo {author} {\bibfnamefont {C.}~\bibnamefont
  {Niclass}}, \bibinfo {author} {\bibfnamefont {C.}~\bibnamefont {Favi}},
  \bibinfo {author} {\bibfnamefont {T.}~\bibnamefont {Kluter}}, \bibinfo
  {author} {\bibfnamefont {M.}~\bibnamefont {Gersbach}}, \ and\ \bibinfo
  {author} {\bibfnamefont {E.}~\bibnamefont {Charbon}},\ }in\ \href {\doibase
  10.1109/ISSCC.2008.4523048} {\emph {\bibinfo {booktitle} {ISSCC, IEEE
  International Solid-State Circuits Conference}}}\ (\bibinfo {organization}
  {IEEE},\ \bibinfo {year} {2008})\ pp.\ \bibinfo {pages} {44--594}\BibitemShut
  {NoStop}%
\bibitem [{\citenamefont {Jakob}\ \emph {et~al.}(2006)\citenamefont {Jakob},
  \citenamefont {Schwarzbacher}, \citenamefont {Hoppe},\ and\ \citenamefont
  {Peters}}]{JAKOB2006}%
  \BibitemOpen
  \bibfield  {author} {\bibinfo {author} {\bibfnamefont {C.}~\bibnamefont
  {Jakob}}, \bibinfo {author} {\bibfnamefont {A.}~\bibnamefont
  {Schwarzbacher}}, \bibinfo {author} {\bibfnamefont {B.}~\bibnamefont
  {Hoppe}}, \ and\ \bibinfo {author} {\bibfnamefont {R.}~\bibnamefont
  {Peters}},\ }in\ \href@noop {} {\emph {\bibinfo {booktitle} {Irish Signals
  and Systems Conference, 2006. IET}}}\ (\bibinfo {organization} {IET},\
  \bibinfo {year} {2006})\ pp.\ \bibinfo {pages} {99--103}\BibitemShut
  {NoStop}%
\bibitem [{\citenamefont {Hoppe}\ \emph {et~al.}(2001)\citenamefont {Hoppe},
  \citenamefont {Meuth}, \citenamefont {Engels},\ and\ \citenamefont
  {Peters}}]{hoppe2001design}%
  \BibitemOpen
  \bibfield  {author} {\bibinfo {author} {\bibfnamefont {B.}~\bibnamefont
  {Hoppe}}, \bibinfo {author} {\bibfnamefont {H.}~\bibnamefont {Meuth}},
  \bibinfo {author} {\bibfnamefont {M.}~\bibnamefont {Engels}}, \ and\ \bibinfo
  {author} {\bibfnamefont {R.}~\bibnamefont {Peters}},\ }in\ \href {\doibase
  10.1049/ip-cds:20010363} {\emph {\bibinfo {booktitle} {IEEE Proceedings
  Circuits, Devices and Systems}}},\ Vol.\ \bibinfo {volume} {148}\ (\bibinfo
  {organization} {IET},\ \bibinfo {year} {2001})\ pp.\ \bibinfo {pages}
  {267--271}\BibitemShut {NoStop}%
\bibitem [{\citenamefont {Magatti}\ and\ \citenamefont
  {Ferri}(2003)}]{MAGATTI2003}%
  \BibitemOpen
  \bibfield  {author} {\bibinfo {author} {\bibfnamefont {D.}~\bibnamefont
  {Magatti}}\ and\ \bibinfo {author} {\bibfnamefont {F.}~\bibnamefont
  {Ferri}},\ }\href {\doibase 10.1063/1.1525876} {\bibfield  {journal}
  {\bibinfo  {journal} {Review of Scientific Instruments}\ }\textbf {\bibinfo
  {volume} {74}},\ \bibinfo {pages} {1135} (\bibinfo {year}
  {2003})}\BibitemShut {NoStop}%
\bibitem [{\citenamefont {Dertinger}\ \emph {et~al.}(2007)\citenamefont
  {Dertinger}, \citenamefont {Pacheco}, \citenamefont {der Hocht Iris~von},
  \citenamefont {Hartmann}, \citenamefont {Gregor},\ and\ \citenamefont
  {Enderlein}}]{DERTINGE2007}%
  \BibitemOpen
  \bibfield  {author} {\bibinfo {author} {\bibfnamefont {T.}~\bibnamefont
  {Dertinger}}, \bibinfo {author} {\bibfnamefont {V.}~\bibnamefont {Pacheco}},
  \bibinfo {author} {\bibnamefont {der Hocht Iris~von}}, \bibinfo {author}
  {\bibfnamefont {R.}~\bibnamefont {Hartmann}}, \bibinfo {author}
  {\bibfnamefont {I.}~\bibnamefont {Gregor}}, \ and\ \bibinfo {author}
  {\bibfnamefont {J.}~\bibnamefont {Enderlein}},\ }\href@noop {} {\bibfield
  {journal} {\bibinfo  {journal} {ChemPhysChem}\ }\textbf {\bibinfo {volume}
  {8}},\ \bibinfo {pages} {433} (\bibinfo {year} {2007})}\BibitemShut {NoStop}%
\bibitem [{\citenamefont {Colyer}\ \emph {et~al.}(2010)\citenamefont {Colyer},
  \citenamefont {Scalia}, \citenamefont {Rech}, \citenamefont {Gulinatti},
  \citenamefont {Ghioni}, \citenamefont {Cova}, \citenamefont {Weiss},\ and\
  \citenamefont {Michalet}}]{Colyer2010}%
  \BibitemOpen
  \bibfield  {author} {\bibinfo {author} {\bibfnamefont {R.~A.}\ \bibnamefont
  {Colyer}}, \bibinfo {author} {\bibfnamefont {G.}~\bibnamefont {Scalia}},
  \bibinfo {author} {\bibfnamefont {I.}~\bibnamefont {Rech}}, \bibinfo {author}
  {\bibfnamefont {A.}~\bibnamefont {Gulinatti}}, \bibinfo {author}
  {\bibfnamefont {M.}~\bibnamefont {Ghioni}}, \bibinfo {author} {\bibfnamefont
  {S.}~\bibnamefont {Cova}}, \bibinfo {author} {\bibfnamefont {S.}~\bibnamefont
  {Weiss}}, \ and\ \bibinfo {author} {\bibfnamefont {X.}~\bibnamefont
  {Michalet}},\ }\href {\doibase 10.1364/BOE.1.001408} {\bibfield  {journal}
  {\bibinfo  {journal} {Biomedical Optics Express}\ }\textbf {\bibinfo {volume}
  {1}},\ \bibinfo {pages} {1408} (\bibinfo {year} {2010})}\BibitemShut
  {NoStop}%
\bibitem [{\citenamefont {Bestvater}\ \emph {et~al.}(2010)\citenamefont
  {Bestvater}, \citenamefont {Seghiri}, \citenamefont {Kang}, \citenamefont
  {Gr\"{o}ner}, \citenamefont {Lee}, \citenamefont {Kang-Bin},\ and\
  \citenamefont {Wachsmuth}}]{Bestvater2010}%
  \BibitemOpen
  \bibfield  {author} {\bibinfo {author} {\bibfnamefont {F.}~\bibnamefont
  {Bestvater}}, \bibinfo {author} {\bibfnamefont {Z.}~\bibnamefont {Seghiri}},
  \bibinfo {author} {\bibfnamefont {M.~S.}\ \bibnamefont {Kang}}, \bibinfo
  {author} {\bibfnamefont {N.}~\bibnamefont {Gr\"{o}ner}}, \bibinfo {author}
  {\bibfnamefont {J.~Y.}\ \bibnamefont {Lee}}, \bibinfo {author} {\bibfnamefont
  {I.}~\bibnamefont {Kang-Bin}}, \ and\ \bibinfo {author} {\bibfnamefont
  {M.}~\bibnamefont {Wachsmuth}},\ }\href@noop {} {\bibfield  {journal}
  {\bibinfo  {journal} {OPTICS EXPRESS}\ }\textbf {\bibinfo {volume} {18}},\
  \bibinfo {pages} {23818} (\bibinfo {year} {2010})}\BibitemShut {NoStop}%
\end{thebibliography}
\end{document}